\documentclass[%
 reprint,
 amsmath,amssymb,
 aps,
]{revtex4-2}
\usepackage[pdftex, pdftitle={Article}, pdfauthor={Author}, bookmarks=true, colorlinks, linkcolor=blue,
urlcolor=blue, citecolor=blue, plainpages=false, pdfpagelabels, final, breaklinks=true]{hyperref}
\usepackage{graphicx}
\usepackage{dcolumn}
\usepackage{bm}

\begin{document}

\title{Simultaneous nonreciprocal and ultra-strong coupling in cavity magnonics}

\author{Chi Zhang$^1$} 
\author{Zhenhui Hao$^1$} 
\author{Yongzhang Shi$^2$}
\author{Changjun Jiang$^1$} 
\author{Xiling Li$^1$}
\author{C. K. Ong$^{1,3}$}
\author{Daqiang Gao$^{1}$}
\author{Guozhi Chai$^{1}$}
    \email[Correspondence email address: ]{chaigzh@lzu.edu.cn}
\affiliation{$^1$Key Laboratory for Magnetism and Magnetic Materials of the Ministry of Education, Lanzhou University, Lanzhou, 730000, People's Republic of China. 
\\$^2$State Key Laboratory of Crystal Materials, School of Physics, Shandong University, Jinan, 250100, People's Republic of China. 
\\$^3$Department of Physics, Xiamen University Malaysia, Jalan Sunsuria, Bandar Sunsuria, 43900, Sepang, Selangor, Malaysia.}

\date{\today} 

\begin{abstract}
We demonstrate the simultaneous realization of nonreciprocal coupling and ultra-strong coupling in cavity magnonics.
By replacing a copper cylinder with a yttrium iron garnet cylinder within the photonic crystal, we achieve an ultra-strong coupling strength of 1.18 GHz and a coupling efficiency of 10.9\%. 
Nonreciprocal microwave transmission emerges within the photonic bandgap, due to the breaking of time-reversal symmetry through the gyromagnetic and Faraday effects. 
This work establishes a foundation for advanced nonreciprocal devices in hybrid cavity magnonic systems, with promising applications in quantum information processing and microwave isolation.

\end{abstract}

\maketitle
\emph{Introduction.}--Over the past few decades, light-matter interactions have been extensively studied.
The strength of interactions is typically described by the coupling strength. When the coupling strength reaches the order of the bare frequency of the system, the system enters the ultra-strong coupling (USC) regime \cite{Science298.5597,RevModPhys.91.025005,Nat.Rev.Phys.1.19,Phys.Rep.1078.1}. 
In this regime, dynamical behaviors of the system undergo significant changes, providing a new platform for quantum information processing and quantum calculation \cite{Nat.Phys.6.772,PhysRevLett.105.196402,PhysRevLett.106.196405,PhysRevX.5.021035,Nat.Phys.13.1}.

As a prominent example of such interactions, cavity magnonics, has emerged as a versatile platform for exploring hybrid systems \cite{sciadv.1501286,SolidStatePhys.69.47,SolidStatePhys.70.1,SolidStatePhys.71.117,Phys.Rep.979.1}. 
In this regime, information is encoded and transmitted through polariton quasiparticles, which arise from the strong coupling between magnons and photons. Since the pioneering work of Soykal and Flatt\'e in 2010, who first proposed the concept of magnon-photon coupling \cite{PhysRevLett.104.077202,PhysRevB.82.104413}, significant progress has been made in both experimental realizations and theoretical developments, including coherent coupling \cite{PhysRevLett.111.127003,PhysRevLett.113.156401,PhysRevLett.123.107702}, dissipative couplings \cite{PhysRevLett.121.137203,PhysRevB.99.134426,PhysRevApplied.22.064036}, indirect couplings \cite{Appl.Phys.Lett.109.152405,PhysRevLett.118.217201,PhysRevLett.121.203601}, long-Distance Coherence \cite{PhysRevLett.131.106702,PhysRevLett.132.206902}, nonreciprocal couplings (NRC)\cite{PhysRevLett.123.127202,PhysRevB.103.184427,Appl.Phys.Lett.119.132403,PhysRevApplied.14.014035,PhysRevApplied.13.044039,Nat.Commun.15.9014}. 

Particularly, NRC plays a crucial role in enabling unidirectional signal transmission\cite{PhysRevLett.123.127202,PhysRevB.103.184427}.
This is of great importance for designing novel isolation devices based on cavity magnonics, as it can prevent interference from reflected signals, thereby enhancing the stability and efficiency of the system.
However, to achieve nonreciprocity over a broader frequency range, it is necessary to maximize the coupling strength. 
In previous studies, nonreciprocal magnon-photon coupling required the introduction of significant external dissipation, achieved through the competition between coherent and dissipative coupling \cite{PhysRevLett.123.127202}. 
However, high dissipation tends to constrain the coupling strength, making it challenging for the system to reach the USC regime. 
As a result, the simultaneous realization of NRC and USC in cavity magnonics has remained a persistent challenge for researchers.
Since Zhang et al. proposed realizing USC at high frequencies by increasing the size of the yttrium iron garnet (YIG) sphere while reducing the microwave cavity dimensions\cite{PhysRevLett.113.156401}, several researchers have successively demonstrated USC at both low \cite{PhysRevApplied.2.054002,PhysRevB.93.144420,PhysRevB.101.214414,PhysRevApplied.20.024039,PhysRevApplied.16.034029} and room temperatures \cite{PhysRevB.107.214423}.
In previous work, we realized the USC in a two-dimensional (2D) photonic crystal (PC) with a YIG cylinder defect at room temperature, achieving the coupling strength of 1.05 GHz and a coupling efficiency of 11.7\% \cite{Appl.Phys.Lett.115.022407}.
In addition, we achieved the nonreciprocal strong coupling in a system composed of an irregular cavity and large-scale YIG wafer by the gyromagnetism and Faraday effect, enabling the control of unidirectional microwave transmission \cite{PhysRevB.103.184427}.
These works provide new insights and methodologies for achieving simultaneous NRC and USC in cavity magnonics.

In this work, we firstly demonstrate the simultaneous realization USC and NRC within the photonic bandgap of a PC.
We achieve an ultra-strong coupling strength of 1.18 GHz, with a coupling efficiency of 10.9\%. 
The breaking of time-reversal symmetry (TRS) in the system, induced by the gyromagnetic and Faraday effects, enables nonreciprocal microwave transmission within the bandgap. 
Specifically, by tuning the strength and direction of the applied magnetic field, microwaves are allowed to propagate preferentially along one direction, while transmission in the opposite direction is significantly suppressed.
This work not only advances the understanding of cavity magnonics in photonic crystals but also provides a foundational framework for developing novel nonreciprocal devices, such as isolators, in hybrid cavity magnonic systems.

\begin{figure}[htbp]
\centering
\includegraphics[width=7.4 cm, height=6 cm]{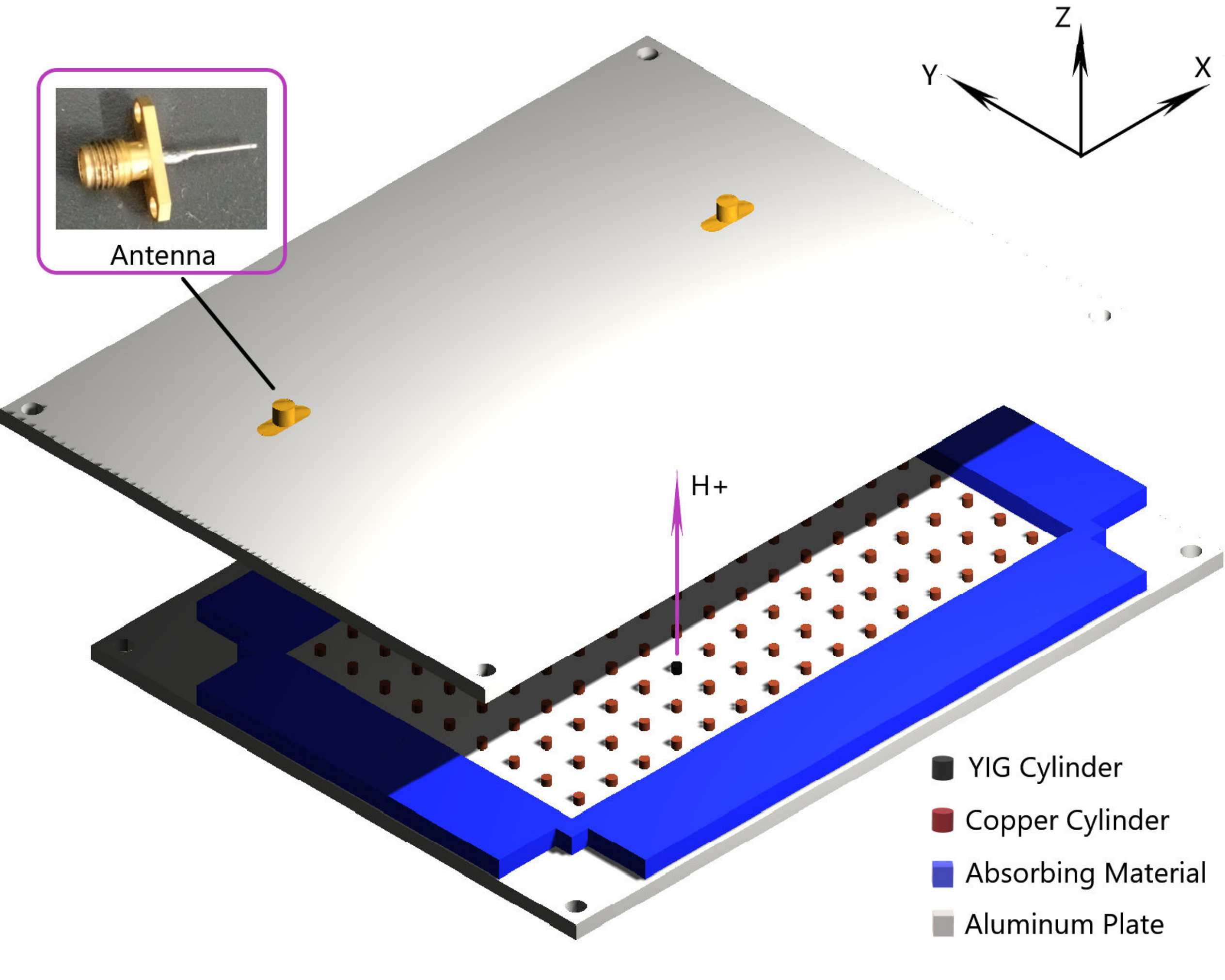}
\caption{Sketch of the experimental setup.
The 2D copper cylinder PC with a point defect of a YIG cylinder.
All cylinders have a diameter of 5 mm and a height of 5 mm.
The lattice constant is 20 mm.
Two aluminum plates are applied to be the two-dimensional chamber with 5 mm of separation and surrounded by some microwave absorbing materials. 
Microwave signals are supplied by a VNA through two microwave cables with two antennas.
The static magnetic field is applied in the $z$ direction.}
\label{Fig:1}
\end{figure}

\emph{Experiment setup}--As illustrated in Fig. \ref{Fig:1}, our device consists of a 2D PC with a point defect by substituting a YIG cylinder for a copper cylinder.
A 2D chamber is constructed by two aluminum plates with 5mm of separation and surrounded by some microwave absorbing materials.
All of the cylinders, which have a diameter and a height as 5 mm are placed in the chamber and the lattice constant of the 2D simple cubic structure is defined as 20 mm.
A vector network analyzer (VNA, Agilent E8363B) is employed to feed microwaves by connecting it to two antennas.
Meanwhile, a static magnetic field along the $z$ direction is applied by an electromagnet.
We choose a single crystal YIG as the magnetic material because of its low microwave magnetic-loss parameter and high-spin-density \cite{J.Phys.D.43.264002}.
Our YIG has a saturation magnetization $M_s$ = 1750 G, a gyromagnetic ratio $\gamma$ = 2.8 MHz/Oe, and a linewidth of 10 Oe.

\begin{figure}[htbp]
\centering
\includegraphics[width=7.4 cm, height=6 cm]{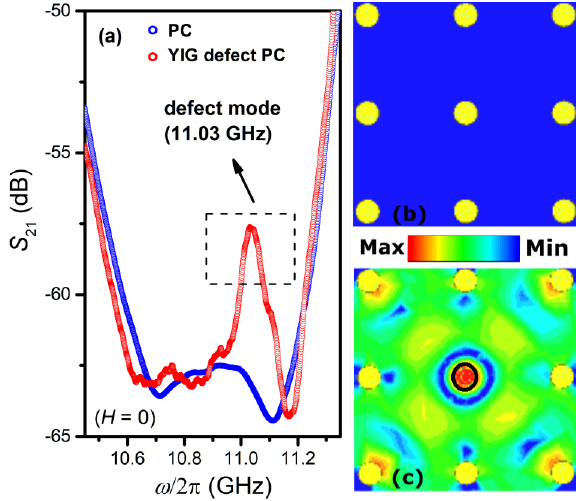}
\caption{(a) Experiment data. Microwave transmission coefficients $S_{21}$ of the PC and YIG defect PC as a function of frequency without applied magnetic field.
(b) and (c) represents the simulation result (by the finite element method) of the strength of microwave magnetic field distribution along $z$ direction of the PC and YIG defect PC at 11.03 GHz, respectively.
Black circle stands for the position of the YIG cylinder.}
\label{Fig:2}
\end{figure}

The scatter parameter $S_{\mathrm{ij}}$ (i, j = 1, 2) is measured to characterize the experimental phenomena discussed subsequently. $S_{\mathrm{ij}}$ represents the microwave transmission signal from port $j$ to port $i$.
Figure \ref{Fig:2}(a) shows the transmission coefficients $S_{21}$ of the copper cylinder PC and the YIG defect PC as a function of frequency.
The microwave magnetic field distribution at 11.03 GHz in the PC and the YIG defect PC is illustrated in Figs. \ref{Fig:2}(b) and \ref{Fig:2}(c), respectively.
It reveals that a strong concentration of the magnetic energy can be observed at the position of the YIG cylinder [black circle on Fig. \ref{Fig:2}(c)].
This implies that a localized mode within the band gap is excited. 
Focusing on this defect mode, we swept the applied static magnetic fields over a frequency range between 10.0 and 11.5 GHz.

\begin{figure*}[htb]
\centering
\includegraphics[width=13 cm, height=8.6 cm]{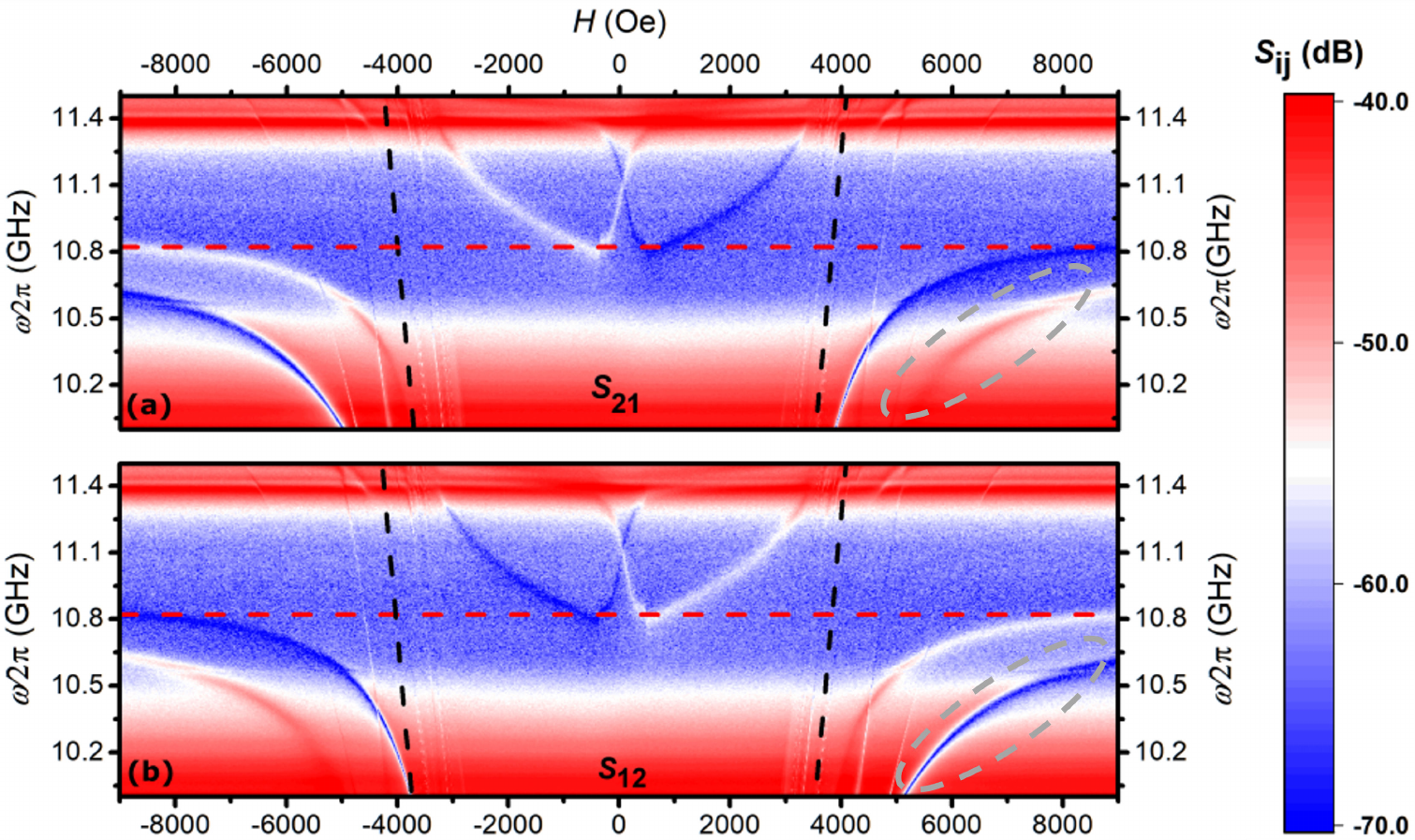}
\caption{Mapping of the amplitude of the microwave transmission as a function of frequency $\omega/2\pi$ and the applied static magnetic field $H$. 
The closer the color to blue, the larger is the microwave transmission loss. 
Black dashed line indicates the FMR mode, calculated with the Kittel equation [Eq. \ref{eq:1}] and red dashed line represents the photon mode. 
(a) $S_{21}$ mapping. 
(b) $S_{12}$ mapping.}
\label{Fig:3}
\end{figure*}

\emph{Nonrecirocal couplings in band gap}--Figures \ref{Fig:3}(a) and \ref{Fig:3}(b) display the magnitude of the microwave transmission coefficients $S_{21}$  and $S_{12}$, respectively. 
The spectra were measured by altering the static magnetic field $H$ from -9000 to 9000 Oe with a step size of 18 Oe.
An anticrossing behavior arises owing to the coupling between magnons and photons.
According to our experiments, magnons are supplied by the FMR mode, and the photons are introduced by the defect mode of the YIG defect PC.
The FMR mode (black dashed line in Fig. \ref{Fig:3}) can be calculated by the Kittel equation \cite{PhysRev.73.155}:
\begin{equation}\label{eq:1}
\omega_{\rm{K}}/2\pi=\gamma\sqrt{(H+(N_x-N_z)M_s)(H+(N_y-N_z)M_s)}.
\end{equation}
Here, $\gamma$ is the gyromagnetic ratio with a value of 2.8 MHz/Oe. 
The demagnetizing factors cannot be calculated analytically, as the demagnetizing field is not uniform in cylindrically shape magnetized bodies. 
We use the experimental results of the demagnetizing factors of the cylinder with the same dimensional ratio from the textbook \cite{Physics.of.Ferromagnetism}. 
Consequently, the demagnetizing factors $N_x$, $N_y$, and $N_z$ are set to values of 0.365, 0.365 and 0.27, respectively. 
It is noteworthy that the photon frequency $\omega_{P}/2\pi$ is 10.8 GHz (red dashed line in Fig. \ref{Fig:3}) rather than 11.03 GHz. 
This is because the magnetic moments within the YIG are not fully aligned under a low applied magnetic field (0-490 Oe), where the YIG has not yet reached saturation magnetization. 
As a result, the interaction between YIG and the defect mode of the PC weakens, preventing the system from reaching the anticrossing condition. 
However, this interaction still induces a nonlinear frequency shift in the defect  mode \cite{PhysRevB.93.144420}, and the asymmetry of the microwave antennas and their positions further disrupts the degeneracy of the defect modes, leading to the splitting of the degenerate modes \cite{Phys.Lett.A.359.1}.  
Moreover, one of the degenerate modes, whose frequency decreases, couples with the FMR mode as the photon mode when the YIG is already up to magnetic saturation.

Unlike the nonreciprocal microwave transmission induced by the cooperative effect of coherent and dissipative coupling in an open cavity-magnon system \cite{PhysRevLett.123.127202}, the nonreciprocity arises from the gyromagnetism effect and the Faraday effect in this work.
As shown in Fig. \ref{Fig:3}, the closer the color to blue, the larger is the microwave transmission loss.

First, we observe that prior to the coupling, when the magnetic field is applied in the positive direction ($z$+), the photon mode manifests as a resonance peak in $S_{12}$ and an anti-resonance dip in $S_{21}$. 
This behavior leads to more efficient transmission of $S_{12}$ when the FMR mode couples with the photon mode, with the coupled polariton modes manifesting as resonance peaks. 
Conversely, the transmission of $S_{21}$ is significantly suppressed, and the coupled polariton mode exhibits an anti-resonance dip.
And then, we simulated the distribution of the microwave magnetic field along the z-direction in the YIG at a frequency of 10.8 GHz under an applied magnetic field of 490 Oe for both the $S_{12}$ and $S_{21}$ directions, using the finite element method.
The results show that energy diverges outward in the $S_{12}$ direction, while it converges inward in the $S_{21}$ direction, which agree well with our experimental results (Simulation details are provided in part \uppercase\expandafter{\romannumeral1} of the Supplemental Material \cite{SM}.).
Furthermore, the direction of the applied magnetic field also significantly influences microwave transmission.
For instance, in $S_{21}$, when the applied magnetic field is negative ($-z$), microwaves in the coupling region transmit more efficiently, whereas transmission is markedly suppressed when the field is positive. Conversely, the results observed in $S_{12}$ exhibit the opposite behavior to those in $S_{21}$.

To better understand the aforementioned phenomena, we present the permeability of the material. 
It is noteworthy that the magnetic anisotropy of YIG cylinder is induced by the applied magnetic field. 
When fully magnetized, the permeability $\mu$ will turn into a second-rank tensor $\boldsymbol{\mu}$ as the microwave magnetic field is perpendicular to $H$. $\boldsymbol{\mu}$ is given by \cite{PhysRevLett.106.093903}:
\begingroup
\renewcommand{\arraystretch}{1.5}
\begin{equation}\label{eq:2}
\boldsymbol{\mu}=\left(\begin{array}{ccc}\mu_{r} & -i \mu_{i} & 0 \\ i \mu_{i} & \mu_{r} & 0 \\ 0 & 0 & 1\end{array}\right),
\end{equation}
\endgroup
where $\mu_{r}= 1+\omega_{K}\omega_{m}/(\omega_{K}^{2}-\omega^{2})$, $\mu_{i} = -\omega\omega_{m}/(\omega_{K}^{2}-\omega^{2})$.
Here, $\omega_{m}/2\pi = \gamma M_{s}$ is the characteristic frequency. 
The off-diagonal element in the permeability tensor is induced by the nonzero applied static magnetic field. It can be concluded that the permeability tensor $\boldsymbol\mu$ is asymmetric ($\mu_{xy} \neq \mu_{yx}$), leading to different effective medium parameters experienced by microwaves propagating in the $S_{21}$ and $S_{12}$ directions.
When the magnetic field is applied along the $+z$ direction, the energy propagating in the $S_{12}$ direction is more efficiently enhanced by the resonance of the YIG, while the microwaves in the $S_{21}$ direction cannot effectively couple back to port 1. 
In this case, TRS of the system is broken \cite{PhysRevLett.74.2662}.
The degree of breakage of TRS is characterized as $u= \mu_{i}/\mu_{r}$ \cite{PhysRevLett.106.093903}. 
For instance, $u$ comes up to a value of 98.9\% as $H$ = 2000 Oe and  $\omega/2\pi$ = 11.0 GHz. 
This implies a near complete breaking of the TRS. 
Furthermore, the isolation ratio (ISO) \cite{PhysRevLett.123.127202} can be calculated with a value of 20 dB in coupling regime from Fig. \ref{Fig:3}, which demonstrates a highly effective unidirectional signal transmission (Calculation details are provided in part \uppercase\expandafter{\romannumeral2} of the Supplemental Material \cite{SM}.).

Additionally, it is important to note that in previous studies of cavity-magnonic, small-scale YIG spheres (with diameters less than or equal to 1 mm) were commonly used as magnetic materials to excite magnons. 
However, in this work, the diameter of the YIG cylinder reaches 5 mm. 
Within the frequency range of 10–11 GHz, the effective wavelength of microwaves passing through the YIG is approximately 7.7–6.7 mm, which is comparable to the size of the YIG.
When microwaves propagate through the YIG, the Faraday rotation angle accumulates linearly with the propagation distance due to the Faraday effect.
The larger the YIG diameter, the more pronounced this cumulative effect becomes. 
This cumulative effect is the core mechanism behind the non-reciprocal microwave transmission: when the rotation direction of the polarization plane aligns with the applied magnetic field, energy is transmitted efficiently; conversely, when the rotation direction opposes the applied magnetic field, a mode mismatch occurs at the ports, leading to energy loss.

In addition, the YIG cylinder is not located on the axis of symmetry of the PC, and the antennas are not placed symmetrically.
These operations are also helpful for increasing the nonreciprocity in the system (Researchers often introduce ferrites into resonant cavities with irregular shapes to break the TRS of the systems in many studies about quantum chaos.) \cite{PhysRevLett.103.064101,PhysRevLett.96.254101, PhysRevLett.123.174101, PhysRevLett.115.026801}. 

\begin{figure}[htbp]
\centering
\includegraphics[width=8 cm]{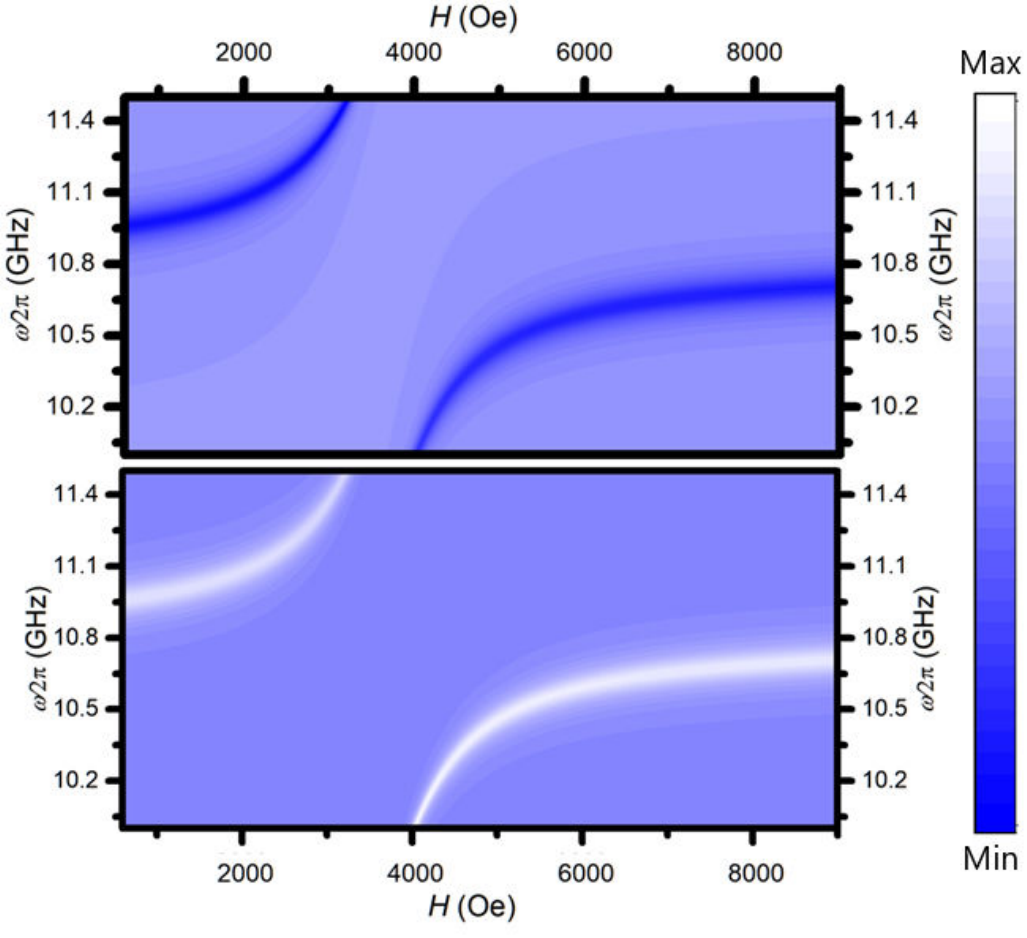}
\caption{Numerical mapping results of the microwave transmission as a function of frequency $\omega/2\pi$ and the applied static magnetic field $H$, calculated with input-output theory[Eq. \ref{eq:3}]. 
The closer the color to blue, the larger is the microwave transmission loss.
(a) $S_{21}$mapping; (b) $S_{12}$mapping.}
\label{Fig:4}
\end{figure}

To better validate the rationality of NRC, we conducted numerical calculations using input-output theory \cite{QuantumOptics}:
\begin{equation}\label{eq:3}
S_{12(21)}=1\pm\frac{\kappa}{i\left(\omega-\omega_c\right)-(\kappa+\beta)+\frac{-(i J+\Gamma)^2}{i\left(\omega-\omega_{\mathrm{m}}\right)-(\alpha+\gamma)}}.
\end{equation}
Here, $J/2\pi$ and $\Gamma/2\pi$ are coherent coupling strength and dissipative coupling strength with a value of 1.18 GHz and 0.01 GHz, respectively.
$\alpha/2\pi$ and $\beta/2\pi$ are the intrinsic damping rate of the magnon mode and photon mode with a value of 0.011 GHz (calculating from the linewidth of YIG cylinder), and 0.1 GHz (fitting from defect mode shown in Fig. \ref{Fig:2}), respectively. 
$\gamma/2\pi$ and $\kappa/2\pi$ are the external damping rates of the magnon mode and photon mode, respectively, where $\gamma=\Gamma^2/\kappa$. 
The numerical results are illustrated in Fig. \ref{Fig:4}, using $J/2\pi$ =1.18 GHz, $\Gamma/2\pi$ = 0.01 GHz, and $\kappa/2\pi$ = 0.1 GHz.
Similarly, the closer the color is to blue, the larger the microwave transmission loss.
It can be observed that the numerical results align well with the experiment results shown in Fig. \ref{Fig:3}.
Moreover, the coherent coupling strength $J/2\pi$ derived from Eq. \ref{eq:3} is significantly greater than the dissipative coupling strength $\Gamma/2\pi$, exceeding it by more than a factor of 100.
This demonstrates that the non-reciprocal coupling observed in this work does not originate from the competition between coherent and dissipative coupling in the system.
The coupling efficiency $\eta$ is determined to be 10.9\% ($\eta=J/\omega_P$, with dissipative coupling strength being negligible), achieving the regime of ultrastrong coupling.

\emph{Ultra-strong coupling}--In order to demonstrate that our system has reached the USC regime theoretically, we conducted a more in-depth analysis of the coupling based on the system's Hamiltonian. 
Typically, the interaction between the cavity photon mode and FMR mode can be described by two coupled harmonic oscillators, with the Hamiltonian given by the Dicke model \cite{NewJ.Phys.21.095004}:
\begin{equation}\label{eq:4}
\hat{H} / \hbar=\widetilde{\omega}_P \hat{c}^{\dagger} \hat{c}+\widetilde{\omega}_K \hat{b}^{\dagger} \hat{b}+\widetilde{g}\left(\hat{c}^{\dagger}+\hat{c}\right)\left(\hat{b}^{\dagger}+\hat{b}\right),
\end{equation}
where $\widetilde{\omega}_P=\omega_{P}+i\beta$, $\widetilde{\omega}_K=\omega_{K}+i\alpha$, and $\widetilde{g}=J+i\Gamma$.
The three terms on the right-hand side of the Eq. \ref{eq:4} represent the photon mode, FMR mode, and the interaction between the two modes, respectively.
In the strong coupling regime, the rotating-wave approximation\cite{PhysRevLett.113.156401} (RWA) can be employed to solve the Hamiltonian, which involves neglecting the counter-rotating terms $\hat{c}^{\dagger}\hat{b}^{\dagger}$ and $\hat{c}\hat{b}$.
This approach has been widely applied in cavity magnonics. However, when the system operates in the USC coupling regime, RWA is no longer valid, and the counter-rotating terms must be taken into account \cite{PhysRevB.72.115303}.

In the USC regime, the Hopfield model is often used instead of the Dicke model to study the system. 
In this case, the Hamiltonian is given by:
\begingroup
\renewcommand{\arraystretch}{1.5}
\begin{equation}\label{eq:5}
\hat{H} / \hbar=\omega_p \hat{c}^{\dagger} \hat{c}+\omega_K \hat{b}^{\dagger} \hat{b}+\widetilde{g}(\hat{c}^{\dagger}+\hat{c})(\hat{b}^{\dagger}+\hat{b})+\widetilde\Lambda(\hat{c}^{\dagger}+\hat{c})^2,
\end{equation}
\endgroup
where $\Lambda=\widetilde{g}^2/2\widetilde{\omega}_K$ is the diamagnetic term, given by Thomas-Reiche-Kuhn sum rule \cite{PhysRevB.101.075301,PhysRevB.107.214423}.
The magnitude of the diamagnetic term directly influences the system's coupling efficiency and phase transition behavior. 
When the diamagnetic term is very small or negligible, the system is more likely to enter the superradiant phase, where the coupling efficiency is significantly enhanced. 
However, in this work, the coupling efficiency has just reached the threshold of the USC regime, and the system has not entered the superradiant phase. 
Therefore, the diamagnetic term cannot be neglected. 
Then, we obtain the eigenvalues by solving the Hamiltonian of the system under these conditions: 
\begin{widetext}
\begin{equation}\label{eq:6}
\begin{aligned}
\widetilde{\omega}_{\pm}=\frac{1}{\sqrt{2}} \sqrt{\widetilde{\omega}_p^2+4 \lambda \widetilde\Lambda \widetilde{\omega}_p+\widetilde{\omega}_p^2 \pm \sqrt{\left(\widetilde{\omega}_p^2+4 \lambda \widetilde\Lambda \widetilde{\omega}_p-\widetilde{\omega}_K^2\right)^2+16 \widetilde{g}^2 \widetilde{\omega}_p \widetilde{\omega}_K}},
\end{aligned}
\end{equation}
\end{widetext}
where $\widetilde{\omega}_{\pm}$ are frequencies of the polariton modes, $\lambda$ is a prefactor added before the diamagnetic term and $g/2\pi$ is coupling strength.

As shown in Figure 5, the blue dots represent the experiment data, while the black curves correspond to the polarized modes $\omega_\pm$ obtained by fitting Eq. \ref{eq:6}. The coherent coupling strength and dissipative coupling strength are fit as $J=1.18$ GHz and $\Gamma=0.01$ GHz, respectively, which are consistent with the results obtained from Eq.  \ref{eq:3}. This confirms that our system has indeed reached the USC regime 
 \cite{Nat.Phys.6.772}. 
 In the ultra-strong coupling region, the stronger the coupling strength, the stronger the nonreciprocity \cite{PhysRevB.103.184427}.

\begin{figure}[htbp]
\centering
\includegraphics[width=7.4 cm, height=7.4 cm]{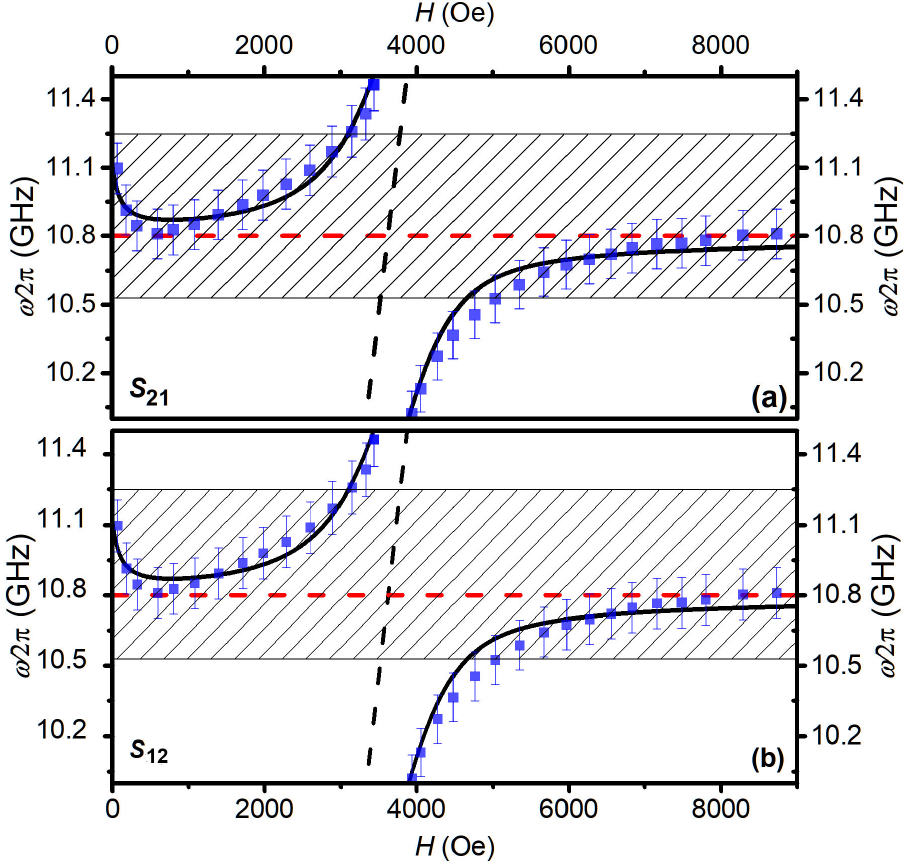}
\caption{The couplings in the frequency versus applied static magnetic field with the fitting curves. (a)$S_{21}$ direction; (b)$S_{12}$ direction. The experiment data are described as blue circles. Black dashed line indicates the FMR mode fitting with the Kittel equation [Eq. \ref{eq:1}];  red dots indicate the photon frequency of defect mode located at 10.8 GHz; green curves are coupling curves fitting with modified Hopfield model  model [Eq. \ref{eq:4}]. Bandgap is marked by dashed gray box.}
\label{Fig:5}
\end{figure}

Such unique dynamic behaviors enable flexible controllability.
In addition, another ultra-strong magnon-photon coupling is observed at the bottom right of Fig. \ref{Fig:3}, marked by the gray dashed ellipses.
We designed another experiment to understand the origin of this coupling and noted that both the magnon mode and the photon mode are provided by the YIG cylinder simultaneously \cite{PhysRevB.91.214430,PhysRevB.93.144420,J.Magn.Magn.Mater.521.167536}.
The coupling strength reaches 4 GHz, and the coupling efficiency is 32.8\%.
Experimental and calculation details are provided in part \uppercase\expandafter{\romannumeral3} of the Supplemental Material \cite{SM}.

\emph{Summary}--In this work, we successfully achieved the simultaneous realization of NRC and USC in the YIG-PC system.
We demonstrated a nonreciprocal microwave transmission within the photonic bandgap by the gyromagnetism and Faraday effect and realized the  USC strength of 1.18 GHz and a coupling efficiency
of 10.9\%.
The breaking of time-reversal symmetry, combined with the system’s asymmetric design, significantly enhanced the nonreciprocal behavior.
Our experimental results, corroborated by theoretical and numerical analyses, confirm that the system operates in the USC regime, where the RWA becomes invalid.

Our work distinguishes itself from previous research by integrating NRC and USC within a single system, offering new insights into cavity magnonics.
Furthermore, it lays the foundation for developing advanced nonreciprocal devices, such as isolators and circulators, in hybrid quantum systems.
Future works will focus on enhancing the coupling strength, improving the quality of microwave transmission, and advancing the miniaturization of principle-based devices.

\section{Acknowledgements}
The authors would like to thank Prof. Peng Yan, Prof. Jinwei Rao, and Dr. Weihua Zhang for useful discussions and suggestions.
This work is supported by the National Natural Science Foundation of China (NSFC) (Nos. 52471200, 12174165 and 52201219).

\bibliography{Ref.bib}
\end{document}